\documentclass[%
preprint,
aps,
amsmath,amssymb,
floatfix,
]{revtex4-2}

\usepackage{amssymb,amsmath,amstext}
\usepackage{graphicx}
\usepackage{bm}
\usepackage{hyperref}
\usepackage{xcolor}

\usepackage{physics}
\usepackage{ulem}
\usepackage{subfigure}
\usepackage{mwe}
\begin{document}

\title{Ferroelectric nematic phase in the system of  perfectly aligned cyllindrically symmetric rods }

\author{Agnieszka Chrzanowska  }
\affiliation{{
Department of Physics, Krak\'{o}w University of Technology,
ul.\ Podchor\c{a}\.{z}ych 1, 30-084 Krak\'{o}w, Poland
\email{a.chrzanowska@pk.edu.pl}
}}
\author{ Lech Longa }
\affiliation{
{{Marian Smoluchowski Institute of Physics, Department of Statistical
Physics and  Mark Kac Center for Complex Systems Research,  Jagiellonian University,
ul. \L{}ojasiewicza 11, 30-348 Krak\'{o}w, Poland}}
\email{lech.longa@uj.edu.pl}
}

\begin{abstract}
The recent experimental discovery of ferroelectric and splay nematic phases
has sparked interest in comprehending the crucial molecular features necessary
to stabilize these innovative structures.  This study advances the ongoing
discourse by investigating the significance of both molecular elongation
and the distribution of molecular dipoles along the main molecular axis.
Using Density Functional Theory, we have established that
a molecular shape characterized by cylindrical symmetry and the presence
of strong parallel dipoles along the symmetry axis
can lead to the self-assembly of a ferroelectric nematic, which is
more stable than the conventional uniaxial nematic phase. Additionally, we provide
criteria for achieving an optimal dipole distribution along the molecular axis.

\end{abstract}

\maketitle

\vspace {0.5cm}

\section{Introduction}

Due to application challenges, liquid crystals (LC) have become a focal point in technological
and scientific research. A particular emphasis is placed on nematic liquid crystals, where no positional long-range
order combines with orientational ordering.  The preferred orientation of this ordering, represented
by a headless vector called the director, can be easily manipulated with an electric field.
Consequently, as light polarization follows this direction, it becomes possible to control
light propagation, laying the fundamentals for liquid crystalline display technology.

Recently, thanks to the efforts of chemists in designing new compounds, a new type of nematic phase
has been experimentally realized, offering a new avenue for technological applications—polar nematic
(or ferroelectric nematic), $N_f$.
While ordinary uniaxial nematics exhibit $D_{\infty h}$
  orientational symmetry, the symmetry of polar nematics is described by
$C_{\infty v}$ symmetry due to the emerging polarization vector field. This field arises when anisotropic molecules contain dipoles that tend to arrange in the preferred direction, typically aligned with the director field. The characteristic textures of these materials are known as furry schlieren textures.
These new LC polar nematics exhibit several advantageous electrical/optical features such as super high dielectric
permittivity, huge spontaneous polarization, strong non-linear optical response, and high electro-optic activity.
Simultaneously, they maintain a high degree of fluidity. Due to these features, they are recognized as special potential
candidates for a wide range of next-generation photonic and electronic devices.

As the first experimentally reported polar nematic arrangement one can mention the work of Hsiung {\em{et al.}} \cite{Hsiung}.
The material studied there is a quasiliquid crystal (QLC), constituting an intrinsic two-component medium composed of spiropyrans in equilibrium with their merocyanine counterparts. The application of the second harmonic generation (SHG) method revealed a noncentrosymmetric arrangement of the ferroelectric type in this system.
The field gained significant traction in 2017 with the synthesis of two LC compounds exhibiting polar order,
$RM734$ and $DIO$ \cite{Mandle, Mandle2, Nishikawa}, earning them a special reputation.
Subsequently, there has been a surge in experimental reports on various physical properties of polar nematics \cite{
Li,Chen2020,Mandle2021,Martelj,MandleMartelj,Mandle2022,Nishi,Nerea,Brown,MandleMartinez,Cruickshank,Connor,Rudquist}.
Other examples of known ferroelectric materials include polymers \cite{Takezoe}, smectics \cite{Novotna, Catalano, Bubnov, Na},
polar chiral nematics \cite{Nishipolarchol, Zhao, Chen}, or systems of ferroelectric nanoparticles immersed in liquid crystals \cite{Garbovskiy}.

A common feature observed in some of these results is that polar nematics  display a certain degree of
spontaneous splay deformation of orientational order. This deformation arises from the wedge shape of molecules,
which organize themselves in Japanese splay stripes, alternating the orientations of narrower edges between the
up and down positions. Within a stripe, the arrangement is locally ferroelectric; however, the overall polarization
is zero, as the stripes are antiferroelectrically arranged. Materials exhibiting this behavior are referred to as
splay polar nematics.

At first   $RM734$ was thought to exhibit splay wave structure of a nanometric size
\cite{Martelj}.
 Now, there are two interpretations in consideration: the splay wave exists on the micro scale \cite{Nerea,MandleMartinez,Connor}, or there is no splay at all \cite{Chen2020,Lavrent}.
%
%
%
From a theoretical standpoint, the existence of polar fluids was envisaged long ago. In the initial theory of liquid crystals proposed as early as 1916, Born \cite{Born}  anticipated the existence of a ferroelectric nematic fluid in which dipoles align in the same direction. While experimental observations of $N_f$ were not made for an extended period, theoretical considerations aimed at explaining how molecular structure underlies the occurrence of the polar nematic phase accumulated over the next almost hundred years. This encompassed models based on the Frank elastic theory, Q tensor Landau model, density functional theory, or Mayer-Saupe-like theory (\cite{Frank, Pleiner, Pleiner1, Meyer1969, Meyer1977, Morozov, Palffy1988, Etxebarria, Manabe, Emel2019, Emelyanenko}).

As previously mentioned, polar nematic phases were initially often associated with spontaneous splay deformations, broken translational symmetry, and spatial inhomogeneity \cite{Frank, Pleiner, Pleiner1}, as a consequence of flexoelectric coupling. This effect arose due to the considered strong geometrical anisotropy of particles with a more pear-like shape, which were large enough to suppress the flipping head-tail process. Such spontaneous splay deformation may offer a means to disrupt polar order. Indeed, findings from \cite{Alberta, Copic_Q, Martelj} indicate that the wedge (or conical-like) shape of particles can lead to a softening of the $K_1$ elastic constant.

Molecular Dynamics simulations based on the Gay-Berne potential \cite{Dhakal, Berardi, Berardi2001} confirm the existence of polar order in tapered particles. Very recently, through the use of Monte Carlo simulations (\cite{Kubala1, Kubala2}), it has also been demonstrated that wedge-shaped rods made from fused hard spheres arrange themselves in splay structures, albeit with density modulations.

Since the initial attempts to explain the polar nematic phenomenon, a recurring question has been whether a polar ferroelectric phase can exist in a system of purely symmetric cylindrical particles endowed with some distribution of electric dipoles. In cases where nonsymmetric shape is deemed necessary, the subsequent question is what type of asymmetry is needed. Another crucial problem is that, so far, the $N_f$ phase has only been observed at high temperatures, making it unsuitable for practical applications. Identifying molecular factors that would promote a ferroelectric nematic phase at lower temperatures is thus another challenge for liquid crystal chemists.

An additional feature observed in dipolar systems is the tendency of molecules to aggregate and form ferroelectric chains. This effect is already present in the case of spherical particles (for instance, see \cite{Sear, Roij, Cailol, Levesque, Dijkstra, Osipov1996}). It is also present in the case of rod-like aspherical particles if the shape aspect ratio is appropriately small \cite{Roij}. This tendency can be a stimulating factor for obtaining polar phases. Indeed, fully atomistic molecular dynamics simulations of polar nematogenic systems predict ferroelectric ordering with a tendency for head-to-tail association of molecules and the creation of polar, chain-like assemblies \cite{Chen, Pelaez, Tiberio, Sims}.

In theoretical considerations of the fundamental properties of nematogenic systems, simple shapes like spherocylinders hold a special place. So far, both in Monte Carlo simulations and density functional theory (DFT) theories \cite{Jackson, Jackson1, Jackson2, Teixeira}, it has been demonstrated that for the case of molecules with a single dipole, no evidence of ferroelectric behavior is observed in any of the states examined, even at low temperatures.

In this paper, we focus on spherocylindrical particles and analyze the possibility of obtaining a polar nematic phase in perfectly aligned rods endowed with one, two, and three dipoles using the ground state analysis and Onsager-type theory.
The assumed cylindrical symmetry of molecules simplifies the analysis, as the primary factors responsible for the potential stabilization of the polar nematic phase should be the distribution of dipoles along the molecule and dipole-dipole interactions, both coupled with the packing induced by the molecular shape. It is also in line with the uniaxial symmetry of $N_f$.
In Section (\ref{Model}), we introduce details of the assumed model. The density functional approach of the Onsager type is presented in Section (\ref{theor}). Results and conclusions are presented in (\ref{resule}). Summing up and discussion are given in Section (\ref{discu}).

\section{Interaction}
\label{Model}

The presence of electric dipoles in molecules is fundamental to the possible existence of a ferroelectric phase. Two dipoles placed parallel to each other (side by side, head-to-head geometry) will repel each other. Conversely, when they are oriented antiparallel (side by side, head-to-tail geometry), they will strongly attract. If hard spherical, rod-like, or other-shaped particles are endowed with a dipole, an excluded region can modify the dipole interactions. This interaction can give rise to a new scenario of particle arrangement, leading to the emergence of completely new phases. This necessitates a special analysis.

The interaction potential
$V$ between two hard-core interacting molecules with dipoles is given by:
\begin{equation}
V= V^{steric}+V^{dd},
\end{equation}
where the first term, $V^{steric}$, accounts for steric interactions depending on the particle's shape,
and the second term, $V^{dd}$, represents the effective interaction term arising from summing up
all the dipole-dipole interactions.
If a particle has $N$ dipoles, the interaction energy follows the superposition rule:
\begin{equation}
V^{dd}=\sum_{i,j} V^{dd}_{ij},
\end{equation}
where $i$ enumerates all the dipoles belonging to the first molecule, and
$j$ pertains to the dipoles of another molecule. $V^{dd}_{ij}$ takes the form
\begin{equation}
V^{dd}_{ij}=\frac{-\mu_i \mu_j}{r'^3_{ij}}
\left[
\frac{3(\vec{s}_i \cdot \vec{r'}_{ij})(\vec{s}_j \cdot \vec{r'}_{ij})}
{\vec{r'}^2_{ij}}-\vec{s}_i \cdot \vec{s}_j
\right], \;\;\; \rm{if \;\;no\;\;overlap},
\label{Vdip}
\end{equation}
where  $\vec{r'}_{ij}$ is the vector joining the dipoles $i$ and $j$,
$\mu_i$ is the strength of the $i$-th dipole moment and $\vec{s}_i$
is the unit vector parallel to the dipole direction.

The formula (\ref{Vdip}) should only be considered when there is no overlap of two particles.
If two particles overlap, and the interaction potential reaches an infinite value,
$V^{steric}=V=\infty$.


\section{Onsager theory of nematic-polar nematic transition }
\label{theor}

The fundamental quantity central to the theoretical framework discussed
in this section is the Mayer function, denoted as $f_{12}$. For the pair potential we have assumed
it takes the form:
\begin{eqnarray} \label{Mayer}
&
f_{12}={\rm{exp}}(-\beta V)-1={\rm{exp}}(-\beta V^{steric}) {\rm{exp}}(-\beta V^{dd})-1=& \\ \nonumber
&\Theta[\zeta(r_{12}-\vec{r}_{12})] {\rm{exp}}(-\beta V^{dd})-1, &
\end{eqnarray}
where  $\zeta(\cdot$) represents the contact distance between two molecules, $\Theta(\cdot)$
 is the Heaviside function, $r_{12}$ is the distance between two molecular centers of mass,  $k_B$ is the Boltzmann constant, $\beta^{-1}=k_B T$, and T is the absolute temperature.

 Even though this function is introduced for the first time in Equation (\ref{excess}),
 we want to highlight a subtlety associated with $f_{12}$ in the context of dipole-dipole interactions.
 Specifically, the integral of $f_{12}$ over the distance vector,
$\vec{r}_{12}$, has dimensions of volume and should be finite. However, for the
assumed interaction, this integral is only conditionally convergent. The standard regularization
procedure involves introducing a convergence factor under the integral:
$f_{12} \longrightarrow e^{-\lambda r_{12}} f_{12}$ and then taking the limit $\lambda\longrightarrow 0$
after performing the integration. An equivalent result is obtained by integrating over spherical shells.
In the subsequent analysis, we will effectively employ the second approach.
%

\subsection{Helmholtz free energy}
\label{free_en}

In studying the possibility of a uniaxial nematic-ferroelectric nematic phase transition, we assume that the nematic order is close to saturation. This allows us to restrict the orientational degrees of freedom of the molecules to two states: the dipoles are parallel or antiparallel according to the geometry, as shown in Fig. (1).
\begin{figure}[htp]
\vspace{0.8cm}
a)\includegraphics[width=0.4\columnwidth]{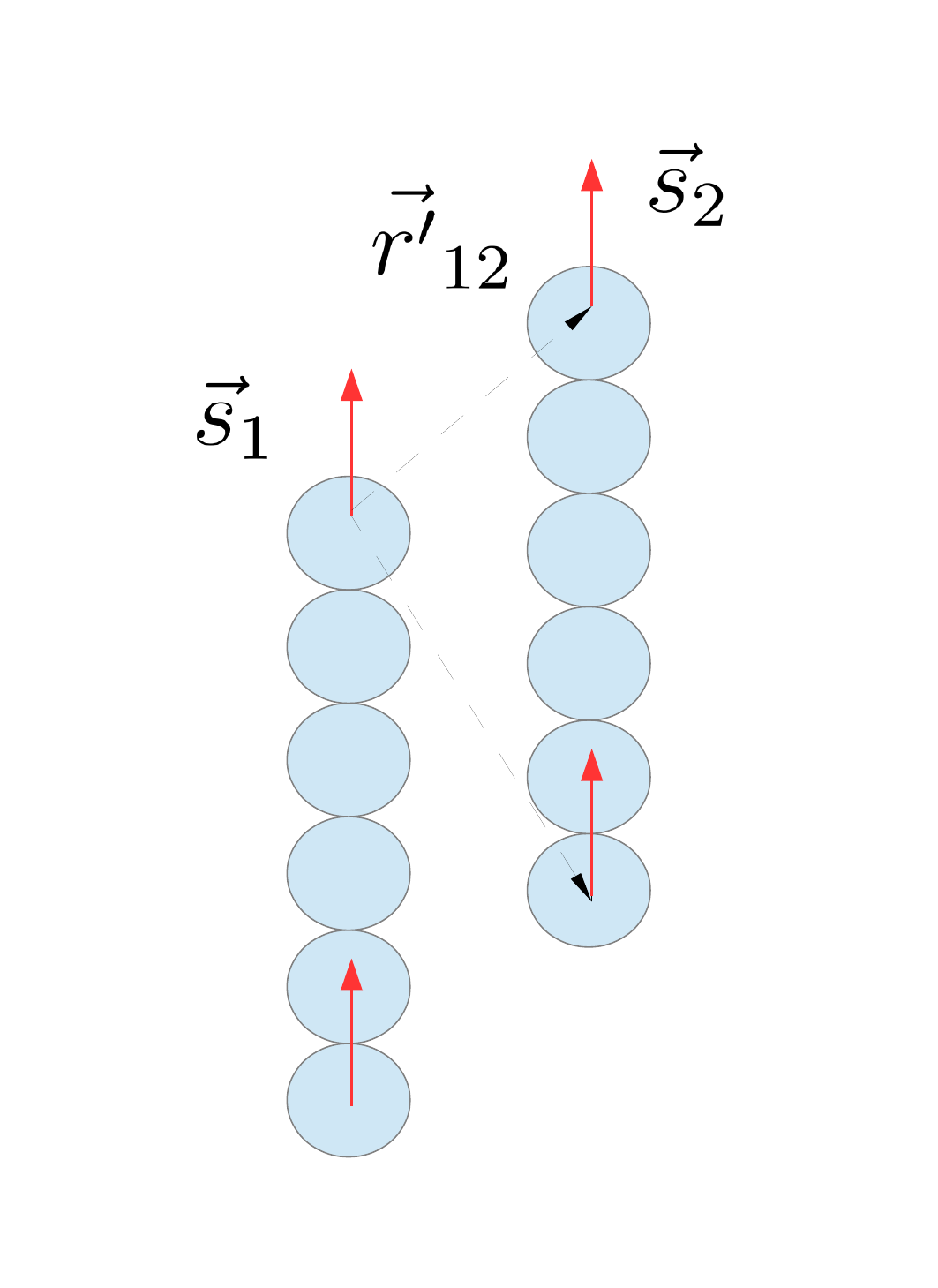}
b)\includegraphics[width=0.45\columnwidth]{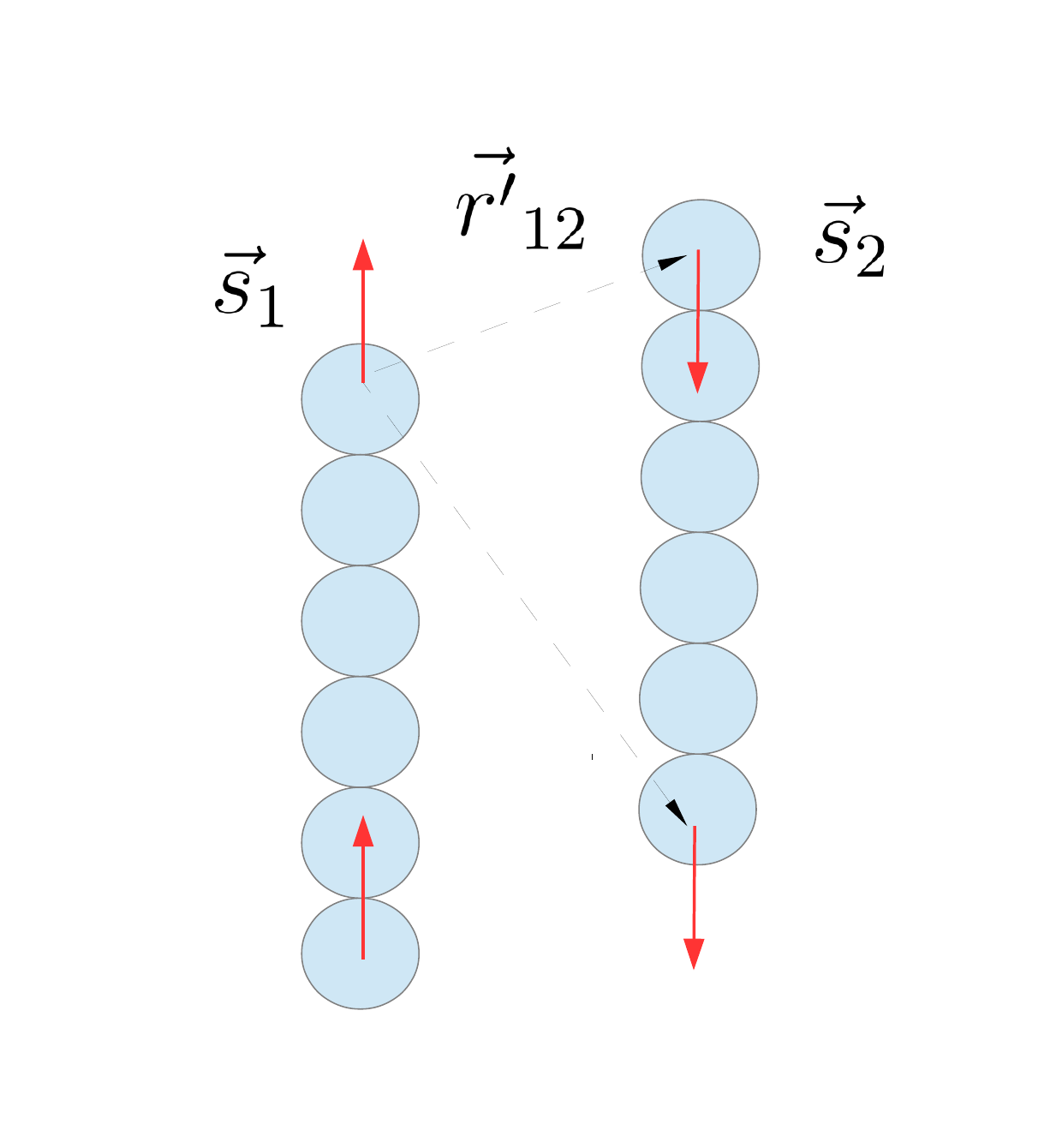}
\caption{Geometry of two interacting hard particles with two dipoles. a) Configuration "up-up" and b) configuration "up-down".}
\label{fig_geom}
\end{figure}
Similar to \cite{Hsiung}, we assume the dipoles to be parallel to the main molecular axis. This choice is supported by the conclusion that in reality, molecules rotate about the longitudinal axis, and even if there are parts of the dipoles transverse to their main molecular axis, their effect is averaged out in the uniaxial phase. It may also happen that the dipoles are nearly longitudinal due to the particle's geometry, as in the case of $RM734$, where DFT calculations show that the effective dipole of the molecule is positioned along the main molecule's axis \cite{Martelj}.
In our approach, the position of the dipole can range from the middle of the particle to the ends of the particle (given by the values of $L_{dipol1}$ and $L_{dipol2}$), excluding the outermost points that would lead to divergence in dipole-dipole interactions.

\begin{figure}[htp]
\vspace{0.8cm}
\includegraphics[width=0.4\columnwidth]{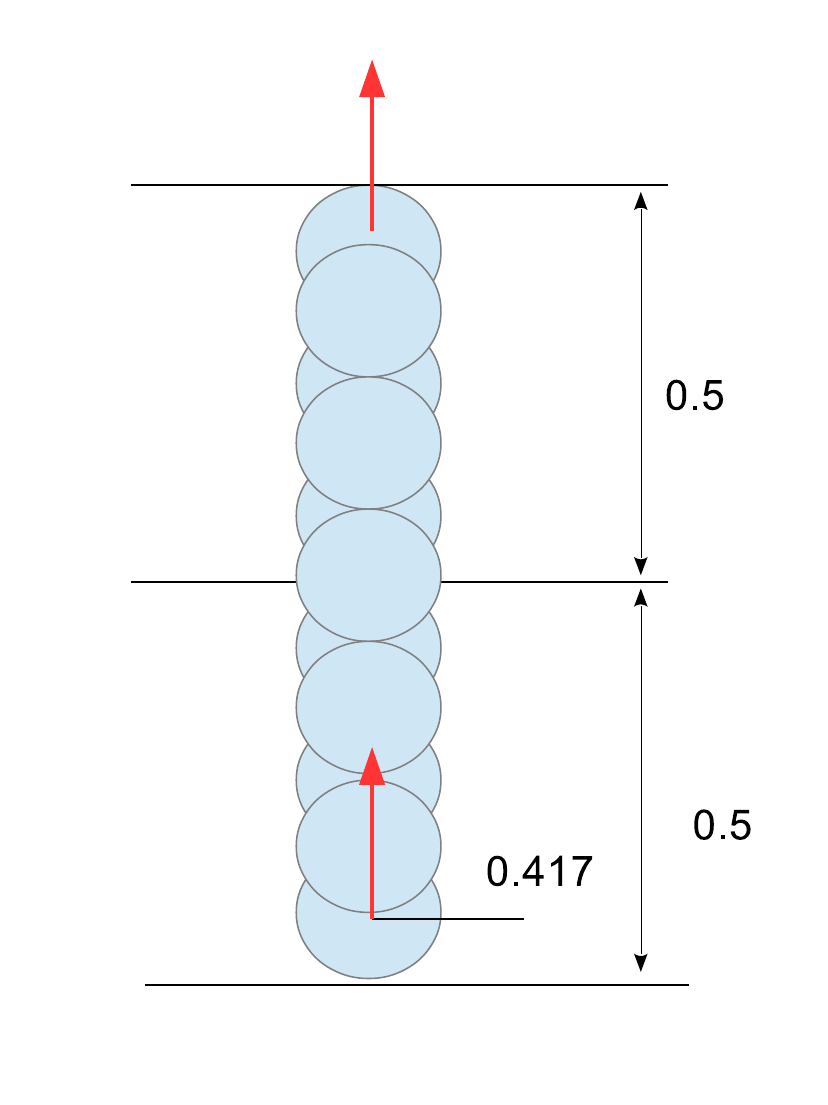}
\caption{Spherocylinder made of spheres. The position of dipoles is counted starting from the particle's center and given with plus sign for upper dipole as well as for lower dipole. $0.417$ is the position of the center of the outermost sphere. }
\label{fig_sfercyl}
\end{figure}

In what follows we will examine two models: one depicting fused spheres,
as illustrated in Fig.(\ref{fig_geom}), and another representing 'spherocylinders', constructed based on spheres,
as shown in Fig.(\ref{fig_sfercyl}).
The length of the molecule is set to 1, and the shape aspect ratio,
denoted as $\kappa$, is defined as the length divided by the breadth, resulting in a value of 6.
The introduced spherocylinder is not an authentic representation of a  spherocylinder shape;
instead, it is a model constructed using 11 spheres where 5 additional spheres are incorporated
into the fused sphere model at the contact points between each pair of spheres. This model proves to be convenient for numerical calculations and effectively replicates the spherocylinder shape.

In the proposed model, under the assumption of perfectly aligned particles, there is only one variable parameterizing molecular orientations. It is denoted as $s$, taking values $s=\pm 1$. The scalar order parameter $<s>$ is associated with an excess in the population of molecules oriented in one direction over the other. Specifically, $s=+1$ represents particles with dipole moments pointing up, while $s=-1$ represents particles with dipole moments pointing down. The order parameter $<s>$ signifies the disparity in the population between molecules oriented "up" and "down". When $<s>$ equals zero, it indicates an equal, on average, distribution of molecules with dipole moments in both directions. Deviations from zero in $<s>$ imply an imbalance in the populations, leading to a perfectly aligned ferroelectric nematic case.
To calculate averages $<>$, one needs the probability distribution function $P(s_i)$, representing the probability of finding a particle $i$ with a given dipole state. Its general form is defined as follows:
\begin{eqnarray}
P(s) = \frac{1}{2}(1+<s>s),
\end{eqnarray}
where $s=\pm 1$.

The nonequilibrium Helmholtz free energy density of the system consists of two terms: the entropic term ${\cal{F}}_0$ and the excess (interaction) term
${\cal{F}}_{ex}$, where
\begin{eqnarray}
{\cal{F}}_0=N \;k_{B} \; T \sum_s P(s) \rm{ln} P(s),
\end{eqnarray}
and where
\begin{eqnarray}
{\cal{F}}_{ex}=-V\;k_{B} \; T
\frac{C(\eta)}{2} \overline{\rho}^2 \sum_{s_1=\pm 1} \sum_{s_2=\pm 1} P(s_1) P(s_2)
\left[
\int d^3 \vec{x}_{12} f_{12} (s_1,s_2,\vec{x}_{12})
\right].
\label{excess}
\end{eqnarray}
Here $V$ is
the volume of the system, $N$ is the number of the particles  and $\overline{\rho}=\frac{N}{V}$
is the average density.
The correcting factor $C(\eta)$  accounts for the neglect of the higher order terms in the  virial expansion.
In the Parsons-Lee approximation  \cite{Parsons,Lee}, which works quite well for hard particles, it follows:
\begin{equation}
C(\eta)=\frac{4-3\eta}{4(1-\eta)^2}.
\end{equation}
Here, $\eta$ denotes the packing fraction:
\begin{eqnarray}
\eta = \frac{N v_0}{V},
\end{eqnarray}
where $v_0$ is the volume of a molecule.

\subsection{Minimisation of the free energy}
\label{mini}

The free energy functional has to be minimized with respect to the distribution function
$P(s)$. This leads to the self-consistency equation for $<s>$
\begin{eqnarray}
<s> =Z^{-1} \sum_{s_1=\pm 1} \rm{exp}^{
C(\eta) \overline{\rho} \sum_{s_{2}=\pm 1}
\overline{f}(s_1 s_2) P(s_2)}
  s_1,
  \label{self}
\end{eqnarray}
where
\begin{eqnarray}
Z= \sum_{s_1=\pm 1}\rm{exp}^{
C(\eta) \overline{\rho} \sum_{s_{2}=\pm 1}
\overline{f}(s_1 s_2) P(s_2)}.
\end{eqnarray}
$\overline{f}(s_1 s_2) $ is the Mayer function averaged over the spatial variable $\vec{r}_{12}$.
It is given by
\begin{eqnarray}
\overline{f}(s_1 s_2) =\int d \vec{r}_{12} f_{12}(\vec{r}_{12},s_1,s_2) = \overline{f}_0+s_1 s_2  \overline{f}_1,
\end{eqnarray}
where
\begin{eqnarray}
 \overline{f}_0 = \frac{1}{4} \sum_{s_1=\pm 1} \sum_{s_2=\pm 1}
 \overline{f} (s_1 s_2) = \frac{1}{2} \left(
 \overline{f}  (+1,+1)  + \overline{f}  (+1,-1)
 \right),
\end{eqnarray}
and where
\begin{eqnarray}
 \overline{f}_1 = \frac{1}{4} \sum_{s_1=\pm 1} \sum_{s_2=\pm 1} s_1 s_2
 \overline{f} (s_1 s_2) = \frac{1}{2} \left(
 \overline{f}  (+1,+1) - \overline{f}  (+1,-1)
 \right).
\end{eqnarray}
Clearly,
$ \overline{f}  (+1,+1)$ is the spatial average of the Mayer function for two particles
in "up up" geometry
and $ \overline{f}  (+1,-1) $ is for "up down" geometry.

 Expanding exponents in (\ref{self}) and considering only linear terms leads to the bifurcation equation:
\begin{eqnarray}
1=C(\eta) \;\;\overline{\rho}\; \;\overline{f}_1,
\end{eqnarray}
which provides the condition for the density
(or packing fraction) at which the bifurcation from the nematic phase to the ferroelectric nematic phase occurs.
The necessary condition is that the density (packing fraction) attains a positive value,
which is the case for $\overline{f}_1 >0$.

As a final comment, let us note that in the numerical calculations,
the forbidden area for particles to approach each other results in an additional
excluded volume term ( the "-1" term in Eq. (\ref{Mayer})) that does not involve
dipolar interactions. However, due to the assumed cylindrical symmetry of the molecules,
this term does not contribute to $\overline{f}_1$ and, hence, has no impact on bifurcation.


\section{Results}
\label{resule}

Before solving Eq. (\ref{self}), the initial step involves analyzing the interaction potential outside the excluded volume for two molecules and determining its global minimum corresponding to $T=0$. If this minimum corresponds to a parallel arrangement of dipoles (geometry "up-up"), it signifies that the polar nematic is the stable solution for $T>0$. An independent bifurcation analysis of the Onsager theory not only validates this result but also furnishes an estimate for the thermodynamic parameters at which the transition from nematic to ferroelectric nematic occurs.

\begin{figure}[htp]
\vspace{0.8cm}
 \includegraphics[width=0.9\columnwidth]{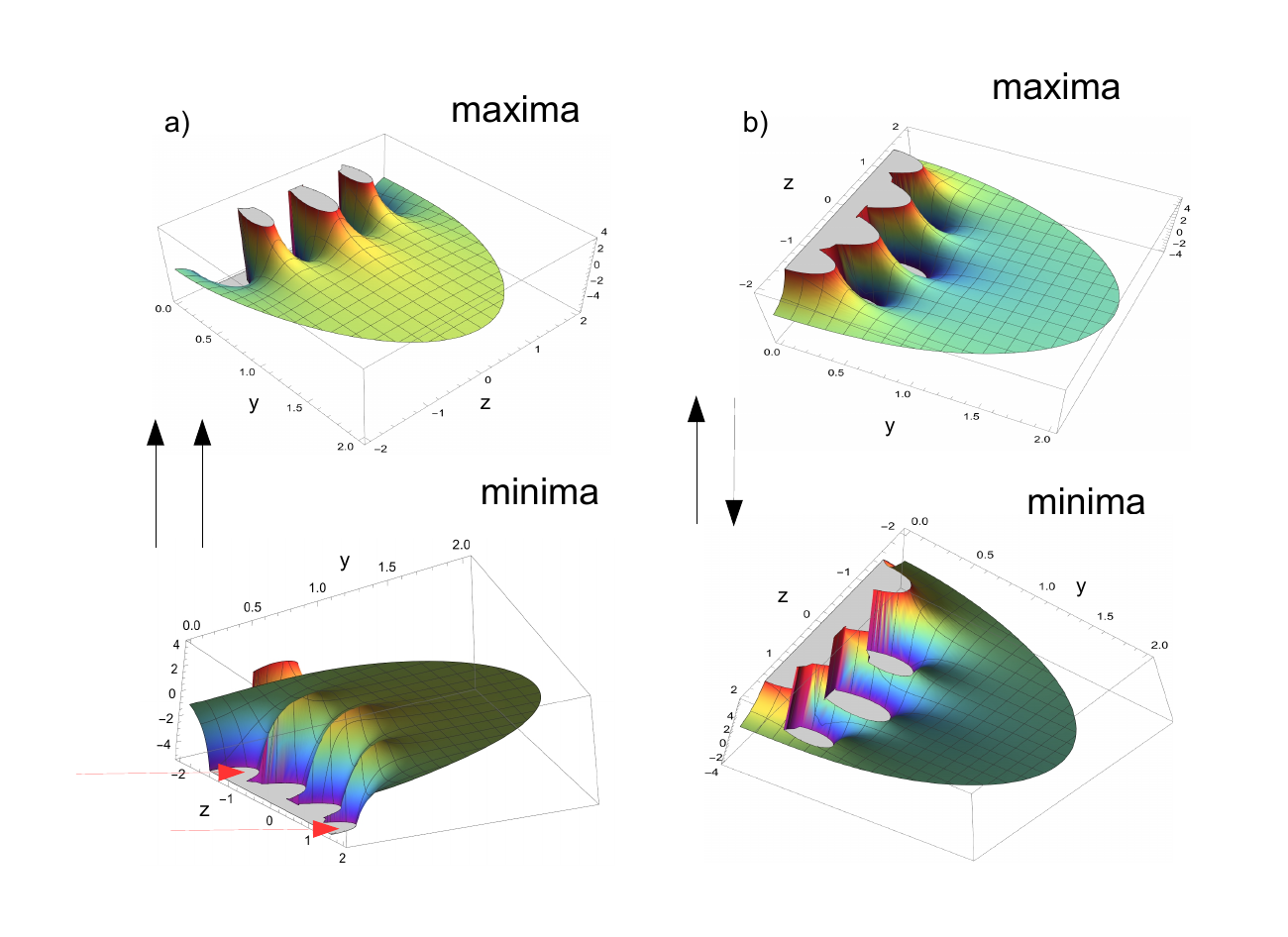}
\caption{ An example of the nontrivial interaction potential shape for the molecules with two dipoles in the  "up up" and "up down" arrangements.   The minima and  maxima are very deep (or high), with the well regions shown in white. The red arrows point towards the deepest minima, occurring for the "up up" arrangement at both sides of the central molecule, specifically at the points $z=+1$ and $z=-1$ (representing the points of the closest approach of the second molecule while the first one is placed at $z=0$). Calculations were performed  using radial coordinates with the condition $r<2$.  }
\label{poteny}
\end{figure}

If the molecular center of particle '1' is at the origin of the reference frame, and its molecular symmetry axis is kept parallel to the $z$-axis, then the pair potential exhibits cylindrical symmetry about the $z$-axis, and the half-plane $\left\{y\ge 0,z \right\}$ can be used for its parameterization. Examples of such parameterized pair potentials for a particle endowed with two dipoles of the same strength are shown in Fig. (\ref{poteny}). Panel a) represents the geometry "up-up," and panel b) represents the geometry "up-down."
In both cases, the minima and maxima occur when the molecular hard cores are in contact. They are very deep and needle-like, respectively, making them challenging to fully illustrate in the scale used in Fig.(\ref{poteny}). The white areas indicate where the potential wells or peaks occur.
In panel a) for "up-up" geometry, the red arrows indicate the deepest minima,
 which occur at both sides of the central molecule at the points $z=+1$ and $z=-1$ (the points of the closest approach of the second molecule while the first one is placed at $z=0$)
suggesting a preference for the polar nematic phase.

\begin{figure}[htp]
\vspace{0.8cm}
\begin{center}
\includegraphics[width=0.7\columnwidth]{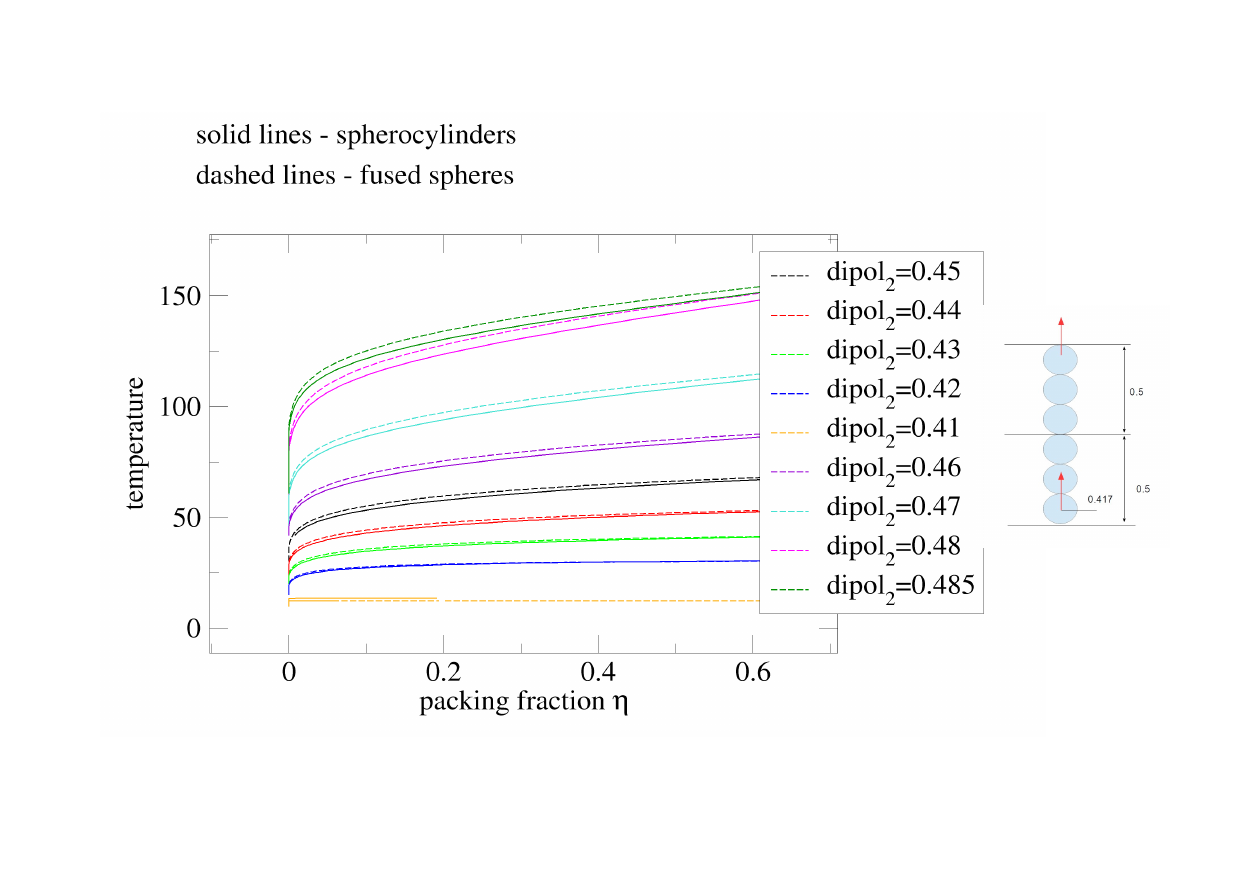}
\end{center}
\caption{ Relation between dimensionless temperatures and packing fractions for the model of fused spheres and cylinders at bifurcation points. The first dipole is positioned at a distance of $L_{\text{dipol1}}=0.43$ from the center of the particle.
The temperature curve is determined by the position of the second dipole, as indicated in the legends.
 }
\label{sfer_cyl}
\end{figure}

Figure (\ref{sfer_cyl}) illustrates the relationship between temperature and packing fraction at the bifurcation point for the model of fused spheres and spherocylinders, both equipped with two dipoles of the same strength. The position of the first dipole is fixed at $L_{\text{dipol1}}=0.43$, determining the distance from the molecule center. The position of the second dipole can be varied from the center of the molecule to its lower end. Interestingly, the ferroelectric phase becomes possible only when the second dipole is placed beyond the value $z>0.41$. There is a clear tendency that as the dipole gets closer to the end of the molecule, the temperature profiles become higher. Larger temperatures of bifurcation can be advantageous for the existence of the polar phase. Although there is a noticeable difference between the two models (fused spheres and spherocylinders), it does not significantly impact the general behavior of temperature profiles.

\begin{figure}[htp]
\vspace{0.8cm}
\begin{center}
\includegraphics[width=0.7\columnwidth]{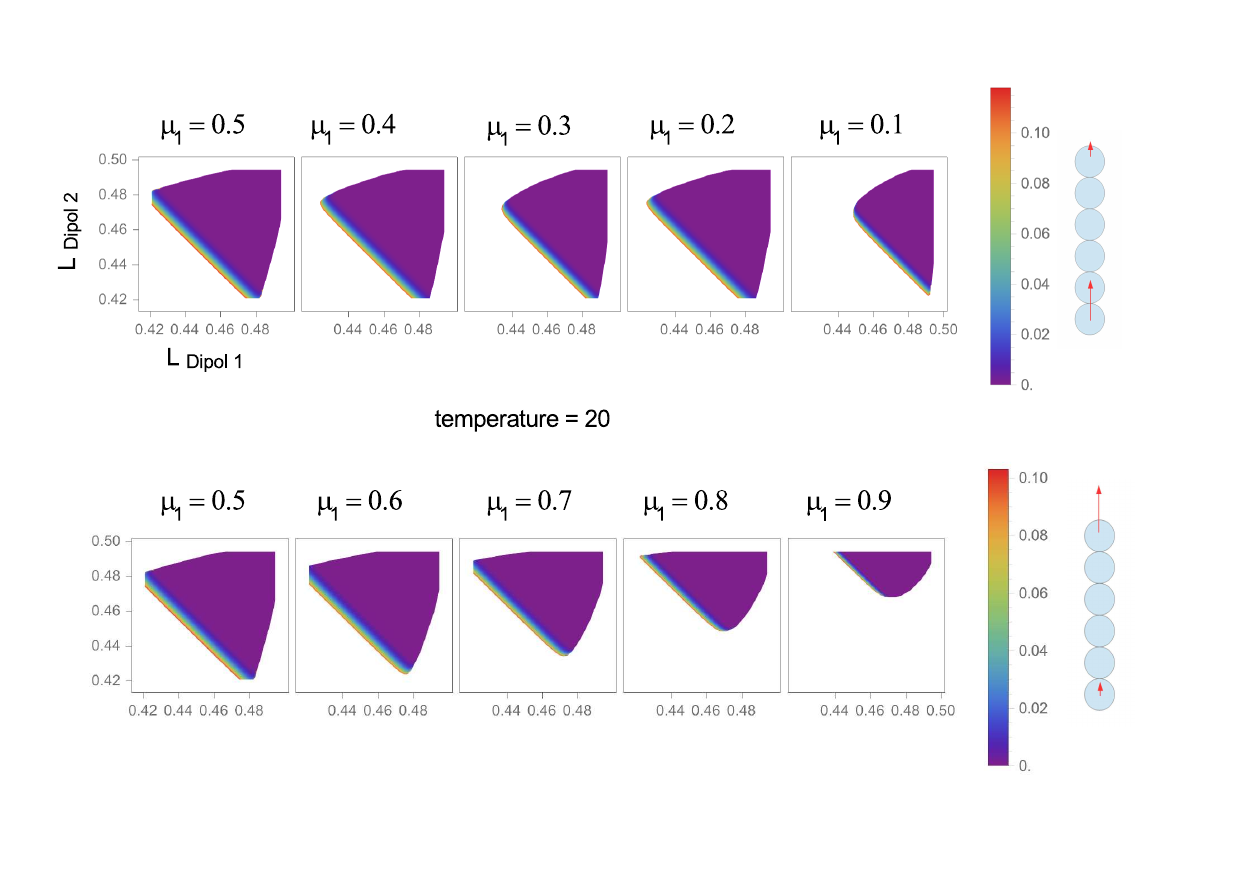}
\end{center}
\caption{The $L_{\text{dipol1}} - L_{\text{dipol2}}$ plane depicts regions where the ferronematic phase can exist for molecules with two dipoles at a temperature of 20. The different colors represent the corresponding packing fractions (degrees specified in the legends). It's important to note that the shape of these regions is influenced by the values of the dipole strengths.
 }
\label{mi_series}
\end{figure}

In Figure (\ref{mi_series}), a plot illustrates the ferronematic regions in the $L_{\text{dipol1}}-L_{\text{dipol2}}$ plane for various dipole strengths. The position of the first dipole, $L_{\text{dipol1}}$, ranges from 0 (the center of the molecule) to 0.5 (the end of the molecule), and the same range is applied to the other dipole position, $L_{\text{dipol2}}$. However, the second dipole is not on the same side as the first dipole. It's worth noting that, due to points close to 0.5 causing one of the integrals to become very large (or infinite for points (0.5, 0.5)), such points are omitted in the drawings.

Each panel in Figure (\ref{mi_series}) is associated with specific values of the dipole strength $\mu$. The largest region is observed when the dipole strengths are equal. Notably, the panels in the upper row differ from those in the lower row. This diference arises from the orientation of the largest dipole; in one row, it points outward from the center of the molecule, while in the other row, it points toward the center of the molecule. It is important that the positions of both dipoles relative to the molecule centers are beyond the center of the outermost spheres used to construct the rods. This might explain why no ferronematic phase was observed in previous theoretical considerations. In most areas of these regions, the packing fractions, apart from narrow stripes of red and yellow colors, are close to zero, indicating the occurrence of the polar phase from the beginning at any density.
Note that the example provided is specific to a chosen temperature (temperature = 20.0), and for other temperatures, these regions will differ, although they occur in similar areas. The condition found for these regions is $L_{dipol1}+L_{dipol2}<\sigma$, where $\sigma$ is the particle breadth or the sphere diameter.

From Figure (\ref{mi_series}), it is evident that when one of the dipoles disappears, there is no possibility to obtain the ferronematic phase. Except for a narrow (colored) stripe, most of the area is violet, indicating a very small packing fraction. This implies that dipole-dipole interactions are so strong that the polar phase exists for any packing fraction, which is a peculiarity of the considered model.

\begin{figure}[htp]
\vspace{0.8cm}
\begin{center}
\includegraphics[width=0.7\columnwidth]{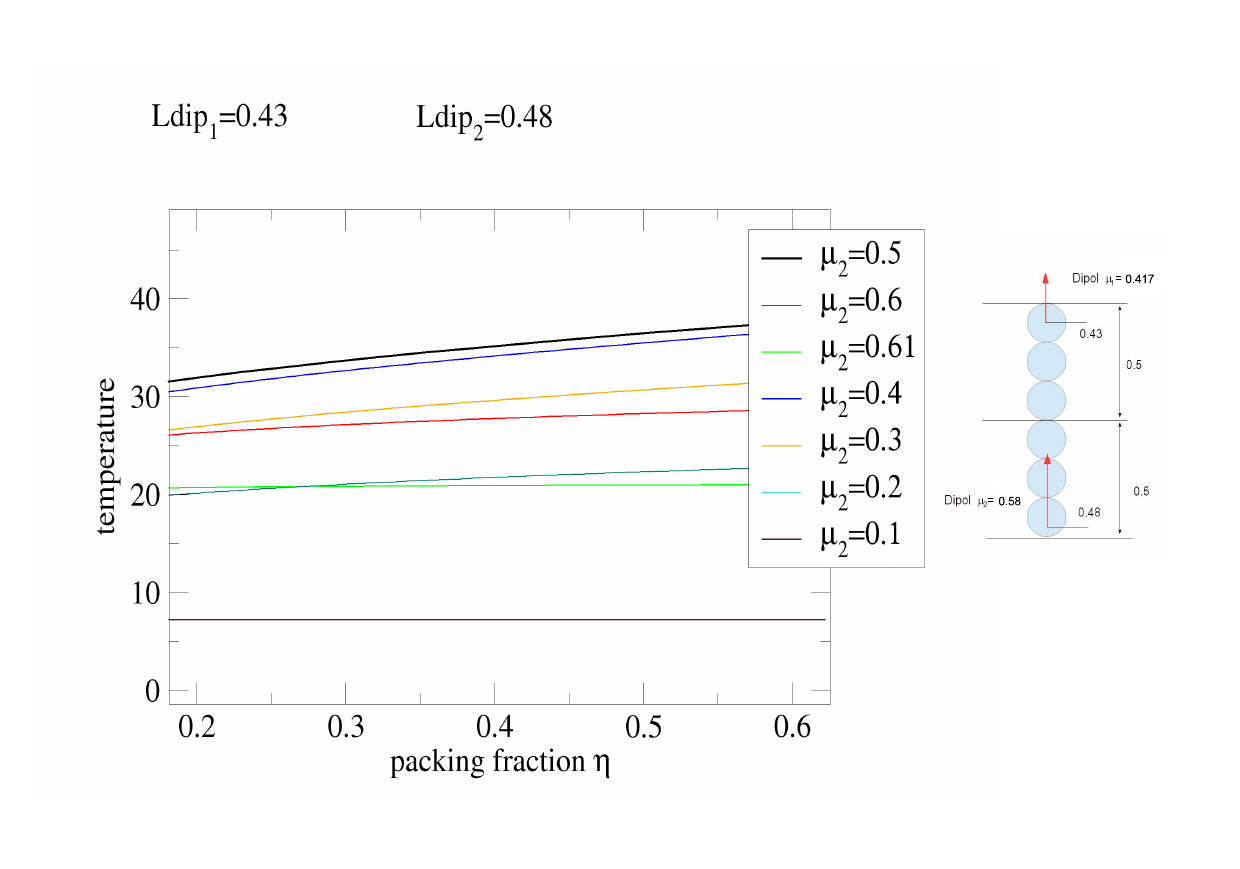}
\end{center}
\caption{Relationship between temperatures and packing fractions at bifurcation points for the model of fused spheres. It provides an example showcasing the influence of dipole strength on these parameters.
 }
\label{pack_mi}
\end{figure}

In Figure (\ref{pack_mi}), we present an example illustrating the influence of dipole strengths on the resulting bifurcation parameters. The first dipole is located at $L_{dipl1}=0.43$, and the second dipole is at $L_{dipol2}=0.48$. The investigation aimed to find optimal values where temperatures would achieve the highest values. It is evident that dipole strengths are most beneficial when they are equal, and this observation was confirmed for various positions of both the first and second dipoles.

\begin{figure}[htp]
\vspace{0.8cm}
\begin{center}
\includegraphics[width=0.7\columnwidth]{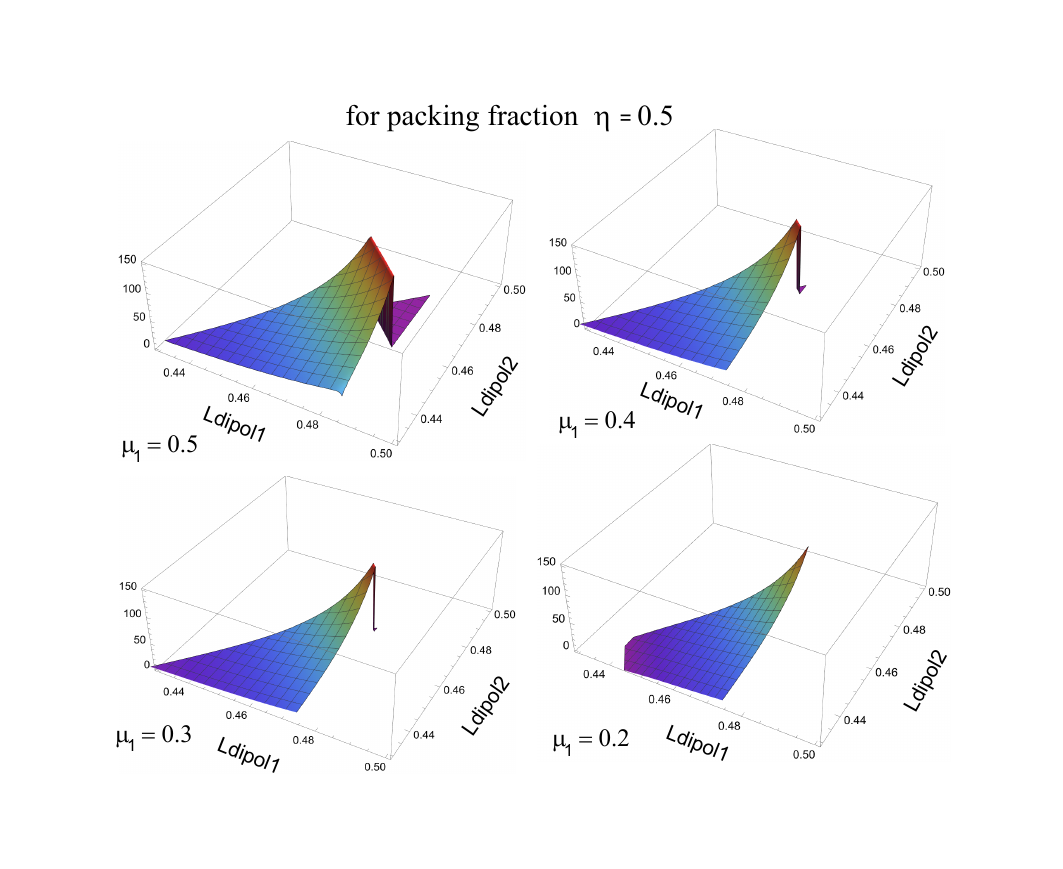}
\end{center}
\caption{ Influence of dipole strengths on temperatures for a fixed packing fraction $\eta=0.5$, while ensuring that $L_{dipol1}>L_{dipol2}$.
 }
\label{pack_mi05}
\end{figure}

It would also be interesting to observe how temperatures depend on the dipole positions for a fixed value of packing fraction. 
In Figure (\ref{pack_mi05}), a packing fraction of $\eta=0.5$ was chosen. The plot reveals that dipole strengths and positions strongly influence the temperature values. Equal strengths of the dipoles consistently yield the highest temperatures for any case where the polar phase occurs, given the applied condition $L_{dipol1}>L_{dipol2}$.

\begin{figure}[htp]
\vspace{0.8cm}
\begin{center}
\includegraphics[width=1.0\columnwidth]{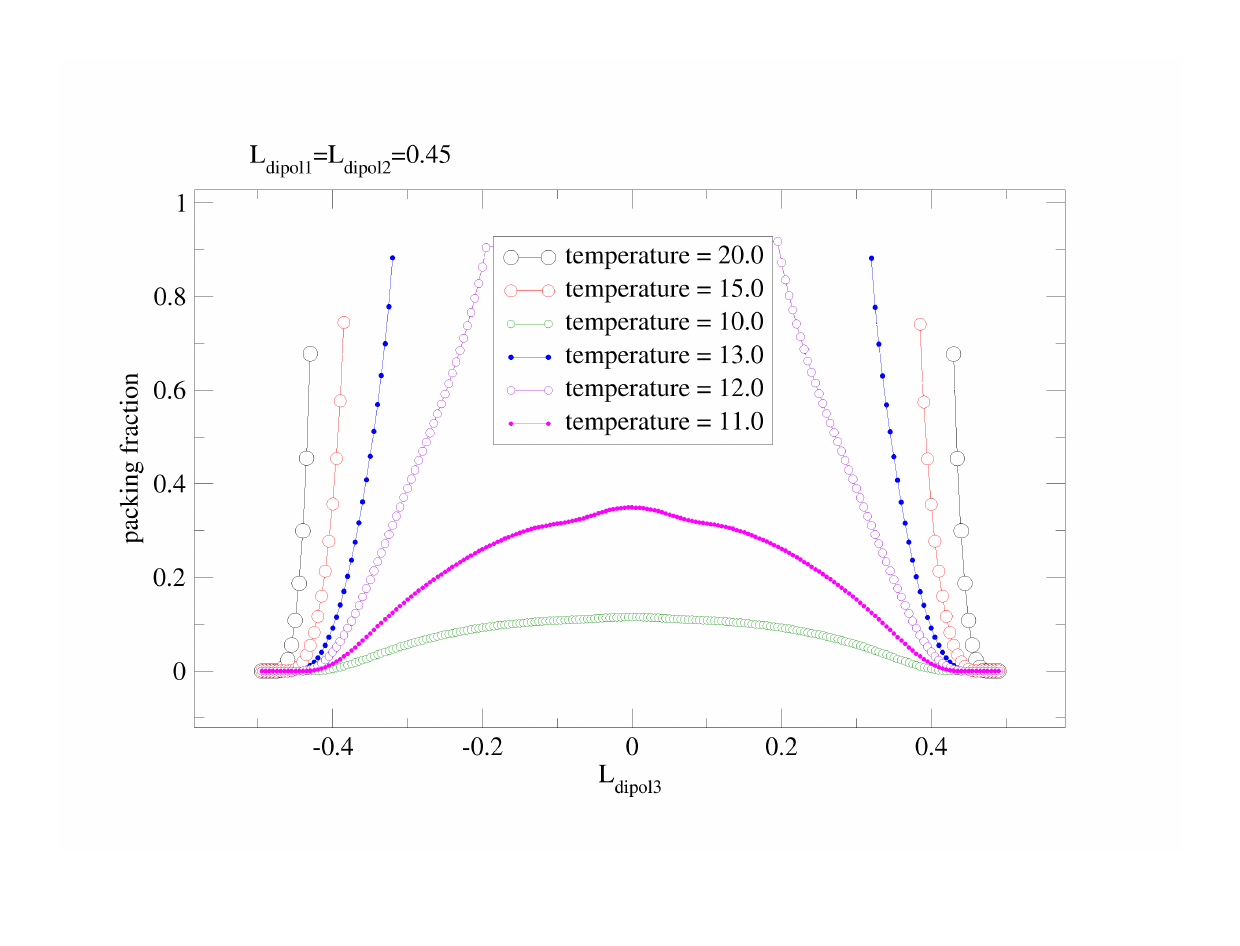}
\end{center}
\caption{ Impact of the third dipole position, $L_{dipol3}$, on the bifurcation parameters
when the first and the second dipole positions are set to $L_{dipol1}=L_{dipol2}=0.45$.
 }
\label{ThirdDip}
\end{figure}

The DFT theory outlined above has also been extended to the system of rod-like fused spheres endowed with three dipoles. In the example provided in Figure (\ref{ThirdDip}), the dipoles used have the same strength ($\mu_i=1/3$) and point in the same direction. It has been observed that the ferroelectric nematic phase is still possible only under the previously identified condition that two dipoles are positioned at the far ends of the molecule. The presence of the third dipole somewhere close to the particle center diminishes the capability of the dipole-dipole interactions at the particle ends, leading to larger values of packing fractions or the disappearance of the polar nematic phase, as observed in the middle of Figure (\ref{ThirdDip}) for temperatures larger than 12.

\section{Summing up}

\label{discu}
In this paper, we have presented and applied a DFT Onsager-type theory for nematogenic systems composed of rod-like particles with simple cyllindrically symmetric shapes and electric dipoles. The focus of the study was to investigate whether the arrangement of one, two, or three dipoles could lead to the emergence of a polar nematic phase. The theoretical results were complemented with a ground-state analysis. From the start we assumed that the ordinary nematic order is saturated, which limits orientational degrees of freedom of a molecule to two dipolar states: up and down.

The Onsager theory proved effective in predicting bifurcation parameters during the transition from ordinary, ideally oriented  nematic to  polar nematic. The theory allowed us to explore dependencies on temperature and packing fractions for different dipole arrangements and strengths. By incorporating the Parsons-Lee scaling idea, more realistic packing fractions could be evaluated.

Several important conclusions were drawn from the study. The most favorable condition for the occurrence of the polar phase was found to be two dipoles positioned at far ends and pointing in the same direction with comparable strength. This dipole arrangement could lead to head-to-tail pairing of molecules, a phenomenon observed in other systems.

The excluded volume of the molecules was identified as an important factor. Experimental observations in previous works, such as \cite{MandleMartinez}, indicated that the existence of lateral parts in molecules could adversely affect the conditions for obtaining polar nematics. Similar effects were observed in our simple model, where the fused spheres exhibited improved conditions for polar phases compared to the spherocylinder shape.

We acknowledge the limitations of our simplified models, each with its own peculiarities. Nevertheless, the conclusions drawn from these models can provide inspiration for further studies and perhaps the development of new real substances. Looking ahead, we raise the question of what geometric shapes or dipole distributions should be considered to accurately mimic real molecules that can form a nematic polar system.

Finally, please note that our current paper assumes perfect alignment in molecules and does not account for bulk lateral parts that might be essential for the creation of nematic splay phases. The consideration of asymmetric molecule models, taking into account lateral components, will be a focus of our future research. This extension will allow us to explore the influence of asymmetry on the formation of different phases and provide a more comprehensive understanding of the behavior of polar nematic systems.

\section{Acknowledgments}
This work was supported by Grant No. DEC-2021/43/B/ST3/03135 of the National Science Centre in Poland.

\end{document}